\documentclass[reprint,eqsecnum,floats,aps,amsmath,amssymb,nofootinbib,prd,onecolumn, showpacs]{revtex4-1}

\usepackage{graphicx}
\usepackage{amsmath,amssymb}
\usepackage{hyperref}
\usepackage{graphicx}
\usepackage{subfigure}
\usepackage{arydshln}
\usepackage{inputenc}
\usepackage{graphicx}
\usepackage{amsmath,amssymb}
\usepackage{hyperref}

\begin{document}

\title{Non-oscillating power spectra in Loop Quantum Cosmology}

\author{Beatriz Elizaga Navascu\'es}
\email{beatriz.b.elizaga@gravity.fau.de}
\affiliation{Institute for Quantum Gravity, Friedrich-Alexander University Erlangen-N{\"u}rnberg, Staudstra{\ss}e 7, 91058 Erlangen, Germany}
\author{Guillermo  A. Mena Marug\'an}
\email{mena@iem.cfmac.csic.es}
\affiliation{Instituto de Estructura de la Materia, IEM-CSIC, Serrano 121, 28006 Madrid, Spain}
\author{Santiago Prado}
\email{santiago.prado@iem.cfmac.csic.es}
\affiliation{Instituto de Estructura de la Materia, IEM-CSIC, Serrano 121, 28006 Madrid, Spain}

\begin{abstract}
We characterize in an analytical way the general conditions that a choice of vacuum state for the cosmological perturbations must satisfy to lead to a power spectrum with no scale-dependent oscillations over time. In particular, we pay special attention to the case of cosmological backgrounds governed by effective Loop Quantum Cosmology and in which the Einsteinian branch after the bounce suffers a pre-inflationary period of deccelerated expansion. This is the case more often studied in the literature because of the physical interest of the resulting predictions. In this context, we argue that non-oscillating power spectra are optimal to gain observational access to those regimes near the bounce where Loop Quantum Cosmology effects are non-negligible. In addition, we show that non-oscillatory spectra can indeed be consistently obtained when the evolution of the perturbations is ruled by the hyperbolic equations derived in the hybrid loop quantization approach. Moreover, in the ultraviolet regime of short wavelength scales we prove that there exists a unique asymptotic expansion of the power spectrum that displays no scale-dependent oscillations over time. This expansion would pick out the natural Poincar\'e and Bunch Davies vacua in Minkowski and de Sitter spacetimes, respectively, and provides an appealing candidate for the choice of a vacuum for the perturbations in Loop Quantum Cosmology based on physical motivations.

\end{abstract}

\pacs{04.60.Pp, 04.62.+v, 98.80.Qc }

\maketitle

\tableofcontents
\clearpage\thispagestyle{empty}

\section{Introduction}

The observational field of high precision cosmology is currently at a full peak of activity, and it is expected to continue growing with feedback from the recent breakthrough of multi-messenger astronomy. The success in this field is especially evident when one takes into account the measurements that have been, and are being performed, of the Cosmic Microwave Background (CMB) \cite{planck,planck-inf}. These observations, when supplemented with a theoretical model, could shed light on quantum gravity effects of the cosmological geometry that may have had an impact on the evolution of primordial fluctuations of the spacetime content. In fact, these fluctuations are believed to constitute the origin of the spatially inhomogeneous degrees of freedom of the Early Universe, and they would have undergone a period of inflation after being originated in epochs of increasingly high energy density and spacetime curvature. In addition, they are believed to be responsible for the measured distribution of temperature anisotropies in the CMB \cite{structures,mukhanov1}. Under such extreme pre-inflationary conditions, it is reasonable to consider that the quantum nature of the cosmological spacetime may have had an important influence on the physical phenomena that took place before inflation.

Over the last decades, there have been several theoretical attempts to incorporate quantum gravity effects in the study of a primordial cosmology with inhomogeneous perturbations (see, e.g., \cite{HalliwellHawking,pintoneto1,pintoneto2, AshLewaDress,Kiefer1,AAN1,AAN2,AAN3,Ivan,hyb-pert1,hyb-pert2,hyb-pert-eff,hyb-pert3,hyb-pert4,hyb-pert5,hybr-ten,hybr-pred,Bojo0,Bojo1,CLB,Bojo2,Edward,Edward2,alesci1,alesci2} and references therein). Furthermore, many of them have succeeded in deriving first approximations to the type of modifications that one can expect to be relevant in the evolution of the perturbations, coming from the quantum behavior of the cosmological spacetime background. In particular, it is worth pointing out interesting investigations on this topic that, within the context of Loop Quantum Cosmology (LQC) \cite{bojo,abl,ashparam,lqcmena}, lead to modifications that are compatible with the observations of the CMB and, at the same time, are potentially capable of capturing  information about the quantum nature of the cosmological background \cite{Ivan,hybr-pred,AshNe}. 

LQC is known to provide a mathematically robust quantization of homogeneous and isotropic cosmologies of the Friedmann-Lema\^itre-Robertson-Walker (FLRW) type, with the remarkable result of generally replacing the cosmological Big Bang singularity with a bounce of quantum origin \cite{APS,mmo}. The theoretical question of how one should include perturbations in the LQC description of FLRW spacetimes has been widely studied over the last years, starting from a variety of different fundamental hypotheses, and with different strategies motivated by first principles and/or phenomenological issues. Among the proposed approaches, let us mention the effective deformed constraint algebra \cite{Bojo0,Bojo1,CLB,Bojo2}, the separate universe framework \cite{Edward,Edward2},  quantum reduced loop gravity \cite{alesci1,alesci2}, the dressed metric formalism \cite{AAN3,AAN1,AAN2,Ivan}, and hybrid LQC \cite{hyb-pert1,hyb-pert2,hyb-pert-eff,hyb-pert3,hyb-pert4,hyb-pert5,hybr-ten,hybr-pred}.  We will focus our attention on the last two strategies. They are two continued lines of research that have led to preliminary predictions that appear to be compatible with cosmological observations at a reasonable level. In hybrid LQC, one considers the cosmological system described in General Relativity by an FLRW metric (typically with compact spatial sections) and an inflaton field with inhomogeneous perturbations, truncates the corresponding Einstein-Hilbert action at lowest non-trivial perturbative order, and identifies a complete set of canonical variables for the description of the FLRW cosmology and the perturbative gauge invariants \cite{hyb-pert4}. The total Hamiltonian of this perturbatively truncated system is then a linear combination of constraints, inherited from those of the Arnowitt-Desser-Misner canonical formulation of full General Relativity. The hybrid strategy to quantize these constraints consists in adopting an LQC representation for the canonical variables that describe the FLRW background, while a more standard Fock representation is employed for the perturbative degrees of freedom. On the other hand, the dressed metric approach to the LQC treatment of cosmological perturbations does not rely on a canonical framework for the entire system. Instead, it starts from the LQC solutions of the FLRW background cosmology, and then lifts the main quantum effects on their dynamics to a physical Hamiltonian for the perturbations \cite{AAN1}. Both theoretical frameworks have been able to provide (effective or mean-field) equations for the gauge invariant perturbations that, while possessing the same local and causal structure as the classical ones, display contributions from the background that contain LQC modifications. In particular, these corrections are able to account for the presence of the bounce of quantum origin that replaces the classical singularity. Furthermore, the state for the homogeneous geometry is often picked out in such a way that these corrections are limited to the vicinity of the bounce, so the evolution of the background cosmology becomes classical very rapidly, and such that the effects of the corrections may be observable today in the CMB \cite{Ivan,hybr-pred}. In such a regime of the quantum theory, the differences between hybrid and dressed metric LQC can be narrowed precisely to a small region around the bounce, and they affect the effective field equations for the perturbations in that region only through the time-dependent mass that appears in them \cite{mass}.

Complete sets of (complex) solutions to the field equations of the gauge invariant perturbations, both for the classical equations and for the quantum corrected ones (whether derived either from hybrid LQC or from the dressed metric approach) give rise to different power spectra that can be eventually confronted with observations. In order to select one of these sets, it suffices to establish initial conditions for the fields at some time of the evolution of the perturbations, thanks to the causal structure of their dynamical equations. Such a choice of initial conditions is usually interpreted as a specification of vacuum state for the perturbations, quantized \`a la Fock. In particular, in the standard cosmological paradigm this initial time is typically set at the onset of inflation, and the data there are fixed to correspond with the Bunch-Davies state \cite{mukhanov1,BD}. This choice is physically reasonable because the standard slow-roll inflationary period is well modelled by de Sitter spacetime, and the Bunch-Davies state is the most natural candidate in this context (i.e., it is the unique Hadamard state that is invariant under the de Sitter isometry group). Remarkably, it provides power spectra for the perturbations that lead to predictions which quite accurately match the observations of the CMB, at least for a large sector of angles in the sky \cite{planck-inf}. However, when the physics that took place before inflation and all the way back to the cosmological singularity (or its quantum analog) is considered to be relevant, the choice of a natural vacuum state for the perturbations fails to be a settled issue. Indeed, the background spacetime in those pre-inflationary epochs, even in the case it remains semiclassical (at least for some stages of the evolution), does no longer resemble de Sitter and its symmetries alone are not enough to fix a unique state. One can restrict this freedom by imposing, on top of invariance of the state under the spatial symmetries, the requirement that the field dynamics is unitarily implementable in the quantum theory (at least in the semiclassical regimes). This criterion actually succeeds in selecting a unique Fock space of states for the perturbations \cite{uniquenessflat,uniquenessrep}; however additional input is needed to single out a preferred vacuum state there.

The choice of a natural state, or equivalently of initial conditions, for the gauge invariant perturbations is a fundamental question to establish the predictive power of any approach to quantum cosmology that provides equations for the perturbations encoding quantum gravity features of the background geometry. In particular, this question needs to be answered if one wants to have any hope of disentangling the possible modifications on the power spectra resulting from genuine quantum cosmology effects from other features of  the spectra caused by alternate choices of vacua, that could also arise in a purely classical pre-inflationary cosmology. Concerning this issue, several proposals have been put forward in the context of hybrid and dressed metric LQC. In these frameworks, a natural choice of initial time to set the data for the perturbations, and thus their vacuum state,  is the moment at which the cosmological bounce happened. Certain low order adiabatic states were first considered owing to their nice properties regarding the renormalizability of the energy-momentum tensor \cite{adiabatic1,adiabatic2,Ivan}. More recently, a somewhat different criterion for the choice of vacuum state has been proposed by Ashtekar and Gupt, based on minimizing the quantum uncertainties of the fields around the bounce and, at the same time, recovering a classical behavior at the onset of inflation \cite{AGvacio1,AGvacio2}. This choice has actually been quite successful in terms of its compatibility with observations in the dressed metric scenario, displaying a slight power suppression for the largest wavelength scales. Nonetheless, all of these vacua lead to power spectra that are highly oscillatory with respect to the scales of observational interest, even in regimes where the pre-inflationary evolution of the background is completely classical, and these oscillations have to be averaged prior to the extraction of predictions. Actually, in the case of adiabatic states, these oscillations are present both in hybrid and dressed metric LQC and they are responsible for an important amplification of the power at medium scales that seems to be in certain tension with the observational data \cite{hybr-pred,Universe}. Even if this possible amplification effect may not be significant in some cases, as it seems to happen with the proposal of Ashtekar and Gupt, one may wonder whether such highly oscillatory behavior of the power spectrum can wash out, or at least obscure, the information about the traces of quantum geometry that the non-Einsteinian evolution of the background cosmology close to the bounce could have imprinted in the dynamics of the perturbations. Motivated by these concerns, Mart\'in de Blas and Olmedo have proposed a different choice of state, based on a selection criterion that is directly tailored to minimize, by numerical methods, the oscillations in the resulting power spectrum \cite{no}. This state has been called the non-oscillating (NO) vacuum state. The resulting power spectrum in hybrid LQC seems to be in very good agreement with observations and, again, predicts power suppresion at large scales.

A theoretical drawback of the two mentioned proposals for the choice of a vacuum for the cosmological perturbations in LQC is that their characterization strongly relies on numerical and/or minimization techniques, that are often interrelated. In this context, the purpose of this work is precisely to provide \emph{analytical} insights supporting a specific characterization of vacuum state, gained by studying some of the general properties of power spectra for gauge invariant perturbations in effective descriptions of LQC that include a period of classical (i.e., Einsteinian) pre-inflationary cosmology. In particular, after a study of the solutions to the field equations for the perturbations using explicitly time-dependent transformations, we provide theoretical arguments that put the focus on power spectra that display NO behavior. Then, starting with the Ermakov-Pinney equation \cite{ermakov,pinney}, we make use of a general formula for the computation of any power spectrum in order to characterize specific conditions that the associated solutions to the field equations must fulfill to minimize the oscillations. After successfully checking that both the Bunch-Davies state in de Sitter spacetime and the Poincar\'e state in Minkowski spacetime satisfy these conditions, we discuss their application to effective regimes of LQC. Finally, we show that, in the ultraviolet regime of short wavelength scales, there is a unique asymptotic expansion of the power spectrum that displays no oscillations at any asymptotic order. This expansion may potentially serve to fix a unique physically privileged vacuum state for the perturbations, and thus a preferred power spectrum, provided that the NO conditions remain satisfied at all scales.

The paper is structured as follows. In Sec. II we formulate the field equation for the perturbations and, analyzing the Hamiltonian that generates this field evolution, we consider time-dependent canonical transformations that render this Hamiltonian diagonal. We use this procedure to construct and conveniently characterize all the normalized solutions. We then provide a qualitative analysis of their power spectra in the context of effective regimes in LQC and argue in favour of the physical importance of finding NO features in it. Sec. III is devoted to the specific characterization of conditions on general power spectra such that they display no scale-dependent oscillations over time, making an auxiliary use of the Ermakov-Pinney equation that is naturally associated with our field equations. We then analyze the feasibility of these conditions in hybrid LQC. In Sec. IV we focus on the ultraviolet sector of short wavelength scales, and perform a study of the oscillatory behavior of the power spectra there. In particular, we show that there is a unique asymptotic expansion for which one can say that no oscillations appear at any order. We end the section remarking on the physical relevance of such expansion in order to fix a natural vacuum state for the perturbations in effective LQC. Finally, in Sec. V we summarize our results. Throughout the paper we work in Planck units, setting $\hbar=c=G=1$.

\section{Solutions from Hamiltonian diagonalization}

Consider a real scalar field $\mathcal{V}(\eta,\vec{x})$, where $\eta$ is a time coordinate and $\vec{x}$ is a triple of spatial coordinates in $\mathbb{R}^3$, with a Fourier expansion in spatial plane waves in which the mode coefficients $v_{\vec{k}}(\eta)$ satisfy the equation
\begin{align}\label{MS}
v_{\vec{k}}''+(k^2+s)v_{\vec{k}}=0,\quad k=|\vec{k}|,\quad \vec{k}\in \mathbb{R}^3-\{0\}.
\end{align}
Here, the primes denote derivatives with respect to $\eta$, and $s$ is a time-dependent real function that we call mass, owing to the formal similarities between this equation and that of a harmonic oscillator with mass. We notice that, for the field $\mathcal{V}(\eta,\vec{x})$ to be real, these mode coefficients must satisfy the reality condition $\bar{v}_{\vec{k}}=v_{-\vec{k}}$. Here and in the following, the bar indicates complex conjugation. Fields with this type of Fourier expansion and dynamics are precisely the ones that describe the gauge invariant perturbations in effective formalisms and mean-field approximations of LQC, when $\eta$ is identified with the conformal time. Specifically, these perturbations are the Mukhanov-Sasaki field for scalar degrees of freedom, and the inhomogeneous contributions of tensor nature to the FLRW metric \cite{MukhanovSasaki,sasaki,sasakikodama}. The power spectra associated with these fields are defined in cosmology as \cite{langlois}
\begin{align}\label{powerv}
\mathcal{P}_{\mathcal{V}}(k,\eta)=\frac{k^3}{2\pi^2}|\mu_k (\eta)|^2,
\end{align}
where $\mu_k$ for all $\vec{k}\neq 0$ is a set of complex solutions to Eq. \eqref{MS} that is required to be normalized according to
\begin{align}\label{normalization}
\mu_k \bar{\mu}_{k}'-\mu_{k}'\bar{\mu}_k=i.
\end{align}
This last requirement on the set of solutions guarantees that the resulting spectra can be directly obtained from the two-point function at equal time of a Fock representation of the field $\mathcal{V}(\eta,\vec{x})$ that is invariant under the Euclidean symmetries of the cosmological background. 

Power spectra are typically evaluated at the end of the (slow-roll) inflationary period in cosmology. Thus, any effect of the dynamical evolution of the perturbations prior to that period that may be observable in the CMB must be found imprinted in the spectra at that moment. If the spectra have oscillated over time during the previous evolution, and these oscillations depend on the Fourier scale, we expect that they will be captured as oscillations in the scale $k$ at the evaluation time. All our following discussions about oscillatory power spectra will keep in mind this relation between the two possible types of dependence of the oscillations. In fact, this very relation is at the heart of the proposal of vacuum state made by  Mart\'in de Blas and Olmedo in Ref. \cite{no}. 

For a general time-dependent mass $s$ any dynamical equation of the form \eqref{MS} can be obtained from the Hamiltonian
\begin{align}\label{MSHam}
H_{\vec{k}}=\frac{1}{2} \left[\big( k^2 + s\big) |v_{\vec{k}}|^2 + |\pi_{v_{\vec{k}}}|^2\right],
\end{align}
where $\pi_{v_{\vec{k}}}$ is to be understood as the canonical momentum of $v_{\vec{k}}$ and satisfies analogous reality conditions. Note that this Hamiltonian generates both the evolution of $v_{\vec{k}}$ and $v_{-\vec{k}}$. In order to study some general properties of the solutions to Eq. \eqref{MS}, starting from this Hamiltonian framework it is convenient to perform explicitly time-dependent canonical transformations of $v_{\vec{k}}$, $\pi_{v_{\vec{k}}}$, and their complex conjugates such that the resulting variables obey Hamilton equations that are purely diagonal. With this purpose, we introduce the transformation
\begin{align}\label{newvar}
a_{\vec{k}}=f_k v_{\vec{k}} + g_k {\bar{\pi}}_{v_{\vec{k}}}, \qquad {\bar{a}}_{\vec{k}}={\bar{f}}_{k} {\bar{v}}_{\vec{k}}+{\bar{g}}_{k}\pi_{v_{\vec{k}}},
\end{align}
where $f_k$ and $g_k$ are unspecified complex functions that depend explicitly on time, and are subject to the constraint
\begin{align}\label{sympl}
f_k {\bar{g}}_{k}-g_k {\bar{f}}_{k}=-i,
\end{align}
that in particular imposes that none of these functions can be zero at any instant of time. Condition \eqref{sympl} is simply the requirement that the introduced transformation is canonical, up to a constant factor $-i$. The Hamiltonian for the new variables $a_{\vec{k}}$, $a_{-\vec{k}}$, and their complex conjugates can be obtained by adding to the former one, given in Eq. \eqref{MSHam}, the explicit time derivative of the generating function of the canonical transformation \eqref{newvar}. The result is
\begin{align}
\tilde{H}_{\vec{k}}=\left[\big(k^2 + s\big)|g_k|^2 + |f_k|^2+\bar{f}_k g_{k}'-\bar{g}_k f_{k}'\right]\left({\bar{a}}_{\vec{k}} a_{\vec{k}}+{\bar{a}}_{\vec{-k}} a_{\vec{-k}}\right) -\left[\big(k^2 + s\big)g_k^2 + f_k^2 -g_k f_{k}'+f_k g_{k}'\right]{\bar{a}}_{\vec k}{\bar{a}}_{-\vec k}+\text{c.c.},
\end{align}
where c.c. indicates the complex conjugate of the preceeding term. This new Hamiltonian generates diagonal equations for $a_{\vec{k}}$, $a_{-\vec{k}}$, and their complex conjugates if and only if
\begin{align}\label{diag}
h_{k}'=k^2+s+h_k^2, \quad  {\rm with}   \quad h_k=f_k g_k^{-1}.
\end{align}
This is an ordinary differential equation of the Riccati type for the function $h_k$, which is equivalent to the set of coupled equations
\begin{align}\label{imre1}
&\text{Re}(h_k)'=k^2+s+\text{Re}(h_k)^2-\text{Im}(h_k)^2, \\\label{imre2}&
\text{Im}(h_k)'=2\text{Re}(h_k)\text{Im}(h_k),
\end{align}
for its real and imaginary parts. Furthermore, one can check that the canonical condition \eqref{sympl} is equivalent to
\begin{align}\label{fgnorm}
 |g_k|^2=-\frac{1}{2\text{Im}(h_k)}.
\end{align}
So, consistency requires that any allowed solution $h_k$ of Eq. \eqref{diag} must have a strictly negative imaginary part. It follows that, given any such complex $h_k$, the resulting diagonal Hamiltonian acquires the form
\begin{align}\label{newham}
\tilde{H}_{\vec{k}}=-\Omega_k\left({\bar{a}}_{\vec{k}} a_{\vec{k}}+{\bar{a}}_{\vec{-k}} a_{\vec{-k}}\right),\quad \Omega_k=F_{k}'+(k^2+s)\frac{\text{Im}(h_k)}{|h_k|^2},
\end{align}
where $F_k$ is the phase of $f_k$. The equations of motion for $a_{\vec{k}}$, $a_{-\vec{k}}$, and their complex conjugates are straightforward to solve in terms of initial data at an arbitrary time $\eta_0$. These, in turn, give rise to solutions of our original equation \eqref{MS}, obtained by simply taking the inverse of the canonical transformation \eqref{newvar}. Specifically, these solutions are
\begin{align}
v_{\vec{k}}=i\bar{g}_{k}e^{-i\int_{\eta_{0}}^{\eta}d\tilde\eta\,\Omega_k(\tilde{\eta})}a_{\vec{k}}(\eta_{0})-ig_ke^{i\int_{\eta_{0}}^{\eta}d\tilde\eta\,\Omega_k(\tilde{\eta})}\bar{a}_{-\vec{k}}(\eta_{0}).
\end{align}
Let us notice that, since $g_k=f_k h_k^{-1}$, each of the two summands in the above solution depends on $F_k$ only through multiplication by the complex exponential of its constant value at $\eta_0$. Furthermore, Eq. \eqref{MS} is linear with real coefficients, so each of the summands in question (multiplied by any constant) provides a complex solution on its own. It follows that we can freely choose $F_k$ as the phase of $h_k$, so that $g_k$ becomes real, and then obtain the following solutions to Eq. \eqref{MS}:
\begin{align}\label{muh}
\mu_k=\frac{1}{\sqrt{-2\text{Im}(h_k)}}e^{i\int d\eta\, \text{Im}(h_k)},
\end{align}
as well as their complex conjugates. It is straightforward to check that these solutions are normalized according to Eq. \eqref{normalization}, which in particular implies that $\mu_k$ and its complex conjugate are linearly independent. Actually, one can see that any complex solution to Eq. \eqref{MS} normalized in this way is of the form \eqref{muh}, with the role of $\text{Im}(h_k)$ played by some strictly negative function which satisfies the same second order differential equation as the imaginary part of $h_k$ [equation that can be derived from Eqs. \eqref{imre1} and \eqref{imre2}] \cite{adiabatic2}. This function is completely fixed once one supplies its value and its first derivative at the initial time $\eta_0$. But we can in fact reproduce any such values by varying the initial data for the real and imaginary parts of $h_k$, in virtue of Eq. \eqref{imre2}. It follows that we can write any normalized solution $\mu_k$ of our original equation \eqref{MS} like in formula \eqref{muh}, where $h_k$ is any solution of the Riccati equation \eqref{diag} with a strictly negative imaginary part. Finally, linear combinations of $\mu_k$ and $\bar{\mu}_k$ provide the general solution to Eq. \eqref{MS}. 

The advantages of characterizing the normalized solutions to Eq. \eqref{MS} by means of formulas \eqref{diag} and \eqref{muh} are many. On the one hand, some general features of these solutions, and of their associated power spectra, can be easily deduced from a direct inspection of the resulting equations for the real and imaginary parts of $h_k$. On the other hand, we will see in Sec. IV that it is possible to characterize a very specific solution to Eq. \eqref{diag} in the asymptotic limit of large $k$ that has an associated spectrum with the most satisfactory properties in this asymptotic regime.

\subsection{NO spectra in effective LQC}

The analysis performed so far is valid for any real function of time $s$, playing the role of a mass in Eq. \eqref{MS}. Let us focus now on the case of cosmological perturbations in the hybrid and dressed metric approaches to inflationary LQC, where the mass $s$ becomes a specific function of the quantum FLRW geometry on the state that describes the background \cite{hyb-pert4,hybr-ten,AAN2,mass}. Inflation is accounted for by the presence of a homogeneous scalar field (with inhomogeneous perturbations), that we call the inflaton, subject to a potential that, for concreteness, we choose to be quadratic in the field. In certain regimes of these LQC models, the power spectrum constructed from a solution to Eq. \eqref{MS} can be understood as the two-point function at equal time of the quantum Heisenberg field operators that describe the Mukhanov-Sasaki perturbations or, as the case may be, the tensor perturbations. Namely, it represents the expectation value on the vacuum state of the product of two field operators, evaluated at different spatial points. This interpretation can be formally justified as follows in the case of hybrid LQC, beginning from the perturbatively truncated system (for specific details and formulas, we refer the reader to Refs. \cite{hyb-pert4,hyb-pert5,fermiLQC,fermiback}). One starts with a specific ansatz for the quantum states in which the wave function factorizes its dependence on the background FLRW geometry and the gauge invariant perturbations, while both parts are allowed to depend on the inflaton field. Searching for states of physical interest, one usually imposes that the partial wave function that describes the FLRW part is close to a solution of homogeneous and isotropic LQC with an inflaton field. Introducing an approximation that is based on the hypothesis of negligible state transitions on the FLRW geometry, the total Hamiltonian can then be reduced to a constraint operator acting only on the partial wave function that corresponds to the gauge invariant perturbations. This hypothesis mathematically amounts to additional conditions on the partial wave function of the FLRW geometry, namely, that it is peaked  with respect to some operators of the homogeneous geometry, that are finite in number (see Ref. \cite{hyb-pert4} for futher details). Remarkably, the resulting constraint on the perturbations depends on the homogeneous geometry only via expectation values of geometric LQC operators on the partial FLRW wave function of the state.  Then, if the Fock representation of the gauge invariant perturbations has been chosen adequately, the Heisenberg evolution of the annihilation and creation operators for the perturbations generated by the aforementioned constraint can be implemented unitarily on Fock space. The construction of the associated unitary operator involves a careful definition of the conformal time to absorb the expectation value of certain geometric operators on the partial state of the FLRW geometry \cite{hyb-pert4,fermiLQC}. In fact, this evolution operator can potentially be used to construct solutions to this constraint, namely (approximate) physical states for the perturbations. Furthermore, the Heisenberg equations deduced in this manner turn out to be precisely of the form of Eq. \eqref{MS}. Our quantum states of interest can then be understood within a context of quantum field theory on a quantum FLRW background, in which the computation of the two-point functions for the gauge invariant fields is equivalent to solving Eq. \eqref{MS} and evaluating the corresponding power spectra. Different solutions to Eq. \eqref{MS} just correspond to different choices of states for the gauge invariant perturbations. It is worth pointing out that this equivalence is completely independent of the specific details of the wave function that describes the cosmological background, provided that  it is close to a solution of homogeneous LQC with negligible geometry transitions, according to our above discussion.  Nonetheless, the details of this FLRW state are transcribed into features of $s$. In what respects the pure computation of the two-point function, in particular, the considered partial FLRW wave function need not obey a semiclassical behavior as long as it fulfills the aforementioned requirements. Moreover, even if the state of the background cosmology does display a prominent semiclassical behavior, the Fock state of the perturbations may still possess genuine quantum features.

Starting from equations of the form \eqref{MS} for the gauge invariant perturbations, where we recall that $s$ contains the most relevant quantum effects of the cosmological background, in this work we will focus only on modifications that are important in what is known as effective LQC. This is a regime obtained by considering a very specific type of state for the background cosmology, motivated as a solution of homogeneous and isotropic LQC, that is peaked on a certain bouncing trajectory. More concretely, this trajectory can be analytically modelled by equations that are of the FLRW type until, torwards the past, the energy density $\rho$ reaches a few percentages of Planck density. For these large densities, the considered equations dictate a departure from General Relativity such that the scale factor $a$ reaches a minimum, corresponding to the moment at which the bounce occurs \cite{ashparam}. This bounce happens at a universal value of the energy density. Furthermore, in order to extract meaningful predictions about the evolution of cosmological perturbations, one typically focuses only on those background trajectories such that the quantum behavior of the Hubble parameter $H$ may only affect the sector of large wavelength scales of the perturbations that are observable nowadays at large angular scales in the CMB. In this way, it is assured that the quantum corrections do not alter the behavior of the shorter scales in the observed spectrum, which is very well explained by General Relativity, while some quantum cosmology modifications may survive in the rest of scales.

The commented solutions for the cosmological background in effective LQC, that are of phenomenological interest for the study of perturbations, are characterized by the following type of initial conditions at the bounce. In what concerns the geometry, on the one hand the Hubble parameter is zero at the bounce, since the scale factor is at a minimum there. On the other hand, with an appropiate use of conformal time, one can make the effective LQC equations (as well as the FLRW ones) only dependent on the relative variation of the scale factor with respect to its value at, e.g., the bounce \cite{continuum}.  We take $a$ as this relative variation; so we have $a=1$ at the bounce. Concerning the homogeneous matter content, our solutions are characterized by an energy density which is dominated by its kinetic contribution at the bounce, namely the contribution from the time derivative of the inflaton, while the potential there is negligible. In the case of a quadratic inflaton potential, it has been proven that these types of initial data are the only ones that lead to LQC effects that may be observable nowadays in the sector of large angular scales of the CMB, while leaving unaffected the rest of scales which are well described by standard FLRW inflationary cosmology \cite{Ivan,hybr-pred,AshNe,Ivan2}. For specific details about the phenomenological viability of the different regions in the parameter space that specifies the initial data of effective LQC (namely, the value of the inflaton at the bounce and its mass), we refer the reader to Ref. \cite{Ivan}. Let us summarize now the typical evolution of such initial conditions, that is well understood after many studies about effective LQC in the literature (see, e.g., the review \cite{Universe}). First of all, the second conformal time derivative of the scale factor, $a''$, is positive at the bounce and, roughly speaking, of a few Planck units in magnitude. This causes that, right after the bounce, a very short superinflationary period occurs. During this period, the rescaled Hubble parameter $aH=a'/a$ grows from zero to a maximum of order one in Planck units, and this happens so fast that the scale factor remains almost constant \cite{AAN3,Universe}. Shortly after the end of the superinflationary period when $H$ reaches its maximum, the quantum corrections to the FLRW equations become completely negligible and the primordial Universe starts a classical phase of deccelerated expansion, according to the dynamics of General Relativity, that is dominated by the kinetic energy of the inflaton. More specifically, the quantum modifications in the evolution become ignorable after the scale factor has increased approximately only $1$ or $2$ e-folds. This small variation is indeed enough to produce a large decrease in the inflaton energy density, to values as small as $10^{-6}$ in Planck units, since its dominant kinetic contribution in this regime is proportional to $a^{-6}$,  while the potential contribution varies very little, with values ranging in the interval $[10^{-12},10^{-11}]$ (see e.g. \cite{Universe,mass}). The classical kinetically dominated phase then goes on until the kinetic and potential contributions to the energy density become comparable, moment at which $aH$ reaches a minimum. After a short transition from the kinetically dominated phase to the domination of the potential, a period of short-lived inflation starts, leading finally to a slow-roll inflationary phase. In fact, the variation of the scale factor from the bounce to the subdominance of the kinetic contribution in the inflaton energy density is typically no more than $4$ or $5$ e-folds. For further details on the semiclassical and classical properties of these background solutions, we refer the reader to Refs. \cite{Universe,Wang1,Wang2,linde,jorma}.

In the evolution equations \eqref{MS} of the Mukhanov-Sasaki and tensor perturbations, the considered effective LQC solutions for the cosmological background are translated into the following features of the mass $s$ in the pre-inflationary period. Since the inflaton potential remains completely negligible from the bounce until almost the end of the kinetically driven classical expansion, in this period we can safely ignore its contribution to $s$ for both types of perturbations. The mass then coincides for the Mukhanov-Sasaki and tensor equations, and is given by \cite{mass}
\begin{align}\label{hdmass}
s=\frac{8\pi}{3}a^2 \rho, \qquad \text{and} \qquad s=-\frac{a''}{a},
\end{align}
for hybrid and dressed metric LQC, respectively, where we recall that the energy density varies in time as $\rho\propto a^{-6}$ in the considered period. The difference between the value of the mass $s$ in the two LQC approaches is due to the fact that, in hybrid LQC, the second time derivative of $a$ is expressed canonically before quantization and then evaluated on effective LQC trajectories, while in the dressed metric approach the scale factor is evaluated at effective trajectories prior to taking its explicit derivatives. These discrepancies can, in turn, be traced back to the strategies followed for the quantization of the perturbations. For more details, we refer the reader to Ref. \cite{mass}. In what concerns this work, the fundamental difference between the two considered masses is their positivity and negativity at the bounce, for hybrid and dressed metric LQC respectively, and the subsequent discrepancies in the superinflationary period. Nonetheless, shortly after the end of superinflation any quantum cosmology correction becomes negligible and both masses coincide there on, in particular during the kinetically dominated classical epoch.

\begin{figure}
	\centering
	\includegraphics[width=13 cm]{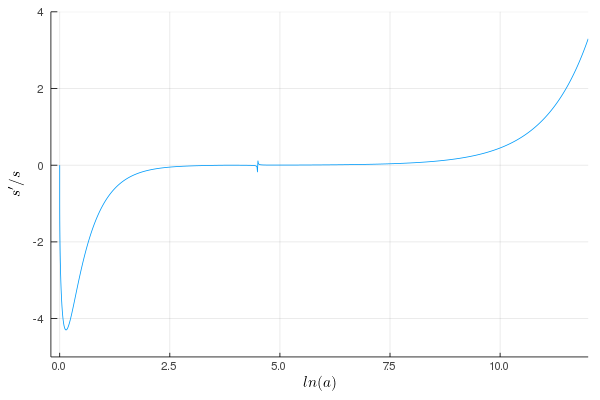}
	\caption{Relative variation of the Mukhanov-Sasaki mass from hybrid LQC in terms of the number of e-folds from the bounce. The plot dispays the Mukhanov-Sasaki mass for the effective FRLW background considered in Ref. \cite{hybr-pred}, determined by an initial value of the inflaton at the bounce equal to $0.97$, and subject to a quadratic potential corresponding to an inflaton mass equal to $1.2\times 10^{-6}$. All quantities are given in Plack units. The (only) apparent discontinuity in the plot shows the moment when this mass becomes negative, close to $4.5$ e-folds.}
	\label{fig1} 
\end{figure}

In Fig. \ref{fig1} we show the relative variation $s'/s$ of the hybrid LQC mass from the bounce to the first epochs of standard slow-roll inflation. Actually, the curve has been computed for the exact expression of the Mukhanov-Sasaki mass, taking into account all contributions from the inflaton potential. However, according to our comments above, this mass must essentially coincide with the one for tensor perturbations approximately until it becomes negative, that is when the potential starts to dominate over the kinetic energy of the inflaton. Indeed, notice that the right-hand side of the first equality in Eq. \eqref{hdmass}, that only takes into account the kinetic contribution, is strictly positive. Furthermore, we recall that the period with relevant LQC effects stops soon after the very rapid super-inflationary stage following the bounce, so the relative variation of the corresponding Mukhanov-Sasaki mass from the dressed metric approach is also given by Fig. \ref{fig1} from a few e-folds on. In particular, even though this mass starts being negative at the bounce  in the dressed metric approach, since $a''>0$ there, it becomes positive during the deccelerated kinetically dominated classical phase, where $a''$ is negative, until the approximate $4.5$ e-folds mark in the figure \cite{mass}.

With this information at hand, let us turn our attention to normalized solutions to Eq. \eqref{MS}, that we have shown that take the form \eqref{muh}, where $h_k$ is a solution of the Riccati equation \eqref{diag} with negative imaginary part, particularized to the case where $s$ is provided by the hybrid  or the dressed metric approaches to LQC. The relevant time-dependent quantity for the computation of power spectra from these solutions is $p_k=|\mu_k|^2$. From Eq. \eqref{imre2} we have
\begin{align}
p_{k}'=\frac{\text{Re}(h_k)}{\text{Im}(h_k)}=-2|g_k|^2\text{Re}(h_k).
\end{align}
It follows that maxima or minima of $p_{k}$ happen at times when the real part of $h_k$ becomes zero and, taking into account Eq. \eqref{imre1}, when
\begin{align}\label{maxmin}
k^2+s-\text{Im}(h_k)^2>0, \qquad \text{or} \qquad k^2+s-\text{Im}(h_k)^2<0,
\end{align}
respectively. In particular, let us notice that, within intervals of time and scales $k$ such that $k^2+s\leq 0$, only one minimum may occur for $p_k$ (if any). Therefore, within such intervals, the power spectra cannot oscillate, regardless of the choice of normalized solutions to Eq. \eqref{MS}. For all other scales and intervals of time where $k^2+s>0$, the power spectrum obtained from normalized solutions may oscillate in time in a way that depends on the scale $k$.

Phenomenologically, using Planck units and setting the reference value of the scale factor at the bounce, the wavenumbers $k$ of the perturbations that had physical (wavelength) scales $a/k$ of the order of the Hubble radius $H^{-1}$ around the bounce in LQC and that are observable today would be in the approximate window $k\in[10^{-1},1]$ (for typical cosmological histories and reheating scenarios). This window can be enlarged with some margin to $[10^{-6},1]$ in order to include the possible observational effects of non-Gaussian correlations with super-Hubble modes \cite{Ivan,IvanG}. These are the scales that should have experienced the most significant LQC effects of the homogeneous geometry. On the other hand,  if we include scales that certainly were not affected by quantum effects, the total window of wavelengths that we can observe nowadays (directly or by non-Gaussian correlations) would approximately be $[10^{-6},10^{2}]$.  Focusing our attention on wavenumbers $k\in[10^{-6},1]$ that might have been influenced by LQC phenomena, we see from Fig. \ref{fig1}, and our discussion about it, that they are such that $k^2+s>0$ in the kinetically dominated classical period after superinflation, both for hybrid and dressed metric LQC. Furthermore, the mass $s$ remains almost constant in this period, with values very close to zero (or numbers much smaller than one). If a power spectrum determined by $p_k$ oscillates in this classical period of the evolution of the background then, for modes $k\ll 1$, the maxima of the oscillations display $\text{Im}(h_k)^2\ll 1$, in virtue of the first inequality in Eq. \eqref{maxmin}. Actually, from Eqs. \eqref{muh} and \eqref{maxmin}, these maxima of $2p_k$ are always greater than $(k^{2}+s)^{-1/2}$ evaluated at those critical points, and therefore much larger than one for $k\ll 1$. It follows that at each instant of time the possible maxima of $2p_k$, viewed as a function of $k$, are bounded from below by the curve $(k^{2}+s)^{-1/2}$, which grows as $k$ decreases, and this curve remains approximately the same throughout the whole interval of time that we are considering, where $s$ varies very little. Similarly, this curve sets an upper bound for the minima of an oscillating $p_k$. In this respect, recall that the bound determined by $(k^{2}+s)^{-1/2}$ reaches orders of magnitude much bigger than one for $k\ll 1$. On the other hand, there is no upper bound of this type for the maxima of $p_k$. These features altogether involve that oscillating power spectra during the kinetically dominated phase, when evaluated later at the onset of inflation, often display a net amplification of the power as $k$ decreases, for those $k$ for which $p_k$ had been oscillating \cite{hybr-pred,Universe,Ivan2}. This statement is also true for power spectra evaluated at the end of inflation in the case of the fields that are often used in cosmology to extract cosmological observations. Indeed, these are not exactly given by the type of field $\mathcal{V}(\eta,\vec{x})$ that we have been considering, but rather by its rescaling with certain functions of the background, that lead to the comoving curvature perturbative field for scalar perturbations, or to its equivalent for tensor ones. The power spectra for these rescaled fields freeze during the period of slow-roll inflation when the wavelength scale $a/k$ gets much larger than $H^{-1}$, so that the analyzed mode crossed the Hubble horizon sufficiently long ago \cite{mukhanov1,langlois}. For the type of solutions that we are studying for the cosmological background, let us recall that the rescaled Hubble parameter $aH$ reaches a minimum at the end of the kinetically dominated phase, with a typical value in the range $[10^{-4},10^{-3}]$ \cite{Universe}. Therefore, we conclude that the modes corresponding to the sector of largest scales under consideration, let us say approximately $k\in[10^{-6},1]$, must have either crossed the horizon well before the onset of inflation or are the first ones to cross after it begins. We then expect that, for such modes, the evaluation of the power spectra of the comoving curvature fields at the end of inflation provides a good picture of the behavior of these fields around the end of the kinetically dominated period and the beginning of the potentially dominated stage, before slow-roll inflation took place.

The presence of oscillations of $p_k$, and their general net effect of amplification of power for small wavenumbers, can be regarded as a somewhat artificial phenomenom if one takes into account that Eq. \eqref{MS} in principle admits normalized solutions such that $p_k$ remains approximately constant during the Einsteinian kinetically dominated phase. Indeed, for this to happen, one would naturally require that
\begin{align}\label{constsol}
\left |\frac{p_{k}'}{p_k}\right |=2|\text{Re}(h_k)|\ll 1,
\end{align}
where we have used Eq. \eqref{imre2}. We can impose this condition at some time $\eta_{i}$ in the beginning of the kinetically dominated phase, and check if it is stable over the whole considered period. For that, it is necessary that we also impose
\begin{align}\label{constsol2}
\text{Im}(h_k)^2(\eta_i)=k^2+s(\eta_i)+r_k, \qquad |r_k|\ll 1
\end{align}
so that $|\text{Re}(h_k)'|\ll 1$ as well in the beginning of the kinetically dominated period. By choosing these initial quantities sufficiently small, the desired condition \eqref{constsol} for an approximately constant spectrum can be made consistent throughout the whole period in question, because Eqs. \eqref{imre1} and \eqref{imre2} imply 
\begin{align}\label{reprimeprime}
\text{Re}(h_k)''=s'+2(k^2+s)\text{Re}(h_k)+2\text{Re}(h_k)^3 -6\text{Re}(h_k)\text{Im}(h_k)^2,
\end{align}
and $s'$ (as well as $s$) remains much smaller than the unit in the kinetically dominated phase. The existence of very slowly varying power spectra during this phase also means that, for the window $k\in[10^{-6},1]$ of modes, the features of the two-point function of the perturbations at the end of the kinetically dominated period can resemble very well those present at the end of the superinflationary period. It is worth recalling that this is the only regime after the bounce where effective LQC corrections are important. One can thus raise doubts as to whether it is physically reasonable to focus the attention on vacuum states for the perturbations that lead to highly oscillating spectra during the times of kinetic domination. Indeed, such oscillatory character can easily erase most of the information coming from the previous epochs near the bounce. Furthermore, these oscillations can result in an enhancement of power that is not due to any quantum cosmology effect nor it is intrinsic to the classical behavior of spacetime after superinflation. Rather, it may correspond to particular features of the specific set of normalized solutions chosen for the perturbations.

In conclusion, we have argued from an analytical perspective that reasonably natural candidates for power spectra which ought to be able to capture most of the genuine LQC corrections on the evolution of the perturbations, without introducing artificial modifications in the part of the pre-inflationary regime that is essentially Einsteinian, are those that present little or no oscillations during such a regime. An argument of continuity of this behavior to the past encourages one to try and characterize NO spectra throughout the entire evolution of the background cosmology, from the bounce to the onset of inflation.

\section{General form of the power spectra: NO conditions}

We have discussed the physical interest of considering vacua, or equivalenty initial conditions, for cosmological perturbations in the framework of effective LQC that lead to power spectra for which the time and $k$-dependent oscillations are minimal. In order to analytically characterize them, we will now study a general formula for the power spectrum associated with any possible vacuum state, conveniently written in terms of a particular solution to the Ermakov-Pinney equation.

Starting from Eq. \eqref{MS} and any set of normalized solutions $\mu_k$, let us call again $p_k=|\mu_k|^2$. This is the function that codifies the freedom of choice of vacuum state in the power spectrum \eqref{powerv}. Using the normalization condition \eqref{normalization}, we have that any such non-zero function $p_k$ satisfies the following second order differential equation
\begin{align}\label{powereq}
p_{k}''+2(k^2+s)p_k=\frac{1}{2p_k}\left[(p_{k}')^2+1\right].
\end{align}
Since $p_k$ is by construction a positive function, we can write it as $p_k=\rho_k^2/2$, where $\rho_k$ is a real non-zero function that, in virtue of Eq. \eqref{powereq}, must satisfy
\begin{align}\label{ermakov}
\rho_{k}''+(k^2+s)\rho_k=\frac{1}{\rho_k^3}.
\end{align}
This is the well-known Ermakov-Pinney equation \cite{ermakov,pinney}. It has been widely employed in the context of FLRW cosmology and its perturbations (see e.g. \cite{it,hawk,kam1,giesel}). Conversely, given any real and non-zero solution $\rho_k$ of this equation, the function $p_k=\rho_k^2/2$ necessarily satisfies Eq. \eqref{powereq}. Therefore, we can completely specify the general solution of this equation if we obtain all possible real solutions $\rho_k$ of the Ermakov-Pinney equation. Actually, this can be done in terms of just one particular solution to Eq. \eqref{ermakov}, in such a way that the resulting formula manifestly displays the possible oscillatory behavior of $p_k$. Let us sketch the procedure to do so. For further details, we refer the reader to Refs. \cite{hawk,milne}.

The general solution of any Ermakov-Pinney equation of the form \eqref{ermakov} can be expressed in terms of two linearly independent solutions to our original equation \eqref{MS}, and their Wronskian \cite{pinney}. These two solutions can, in turn, be chosen as two linearly independent functions, given by one particular real solution $\psi_k$ to Eq. \eqref{ermakov}, multiplied by a sinusoidal function (a sine or a cosine, respectively, for the two considered solutions) of an arc $\phi_k$ such that $\phi_{k}'=\psi_k^{-2}$. Using them, the general real solution of the Ermakov-Pinney equation can be written as
\begin{align}
\rho_k^2=\psi_k^2\left[A\cos^2(\phi_k) +B\sin^2(\phi_k) +C\sin(2\phi_k)\right], \quad C^2=AB-1,
\end{align}
where $A$, $B$, and $C$ are constants that must be real and such that $\rho_k^2$ be positive. Therefore, the function $p_k$ that determines the form of any power spectrum associated with any set of normalized solutions to Eq. \eqref{MS} can be obtained as
\begin{align}\label{generalp}
p_k=\frac{1}{4}\psi_k^2\left[A+B+ (A-B)\cos(2\phi_k)+2C\sin(2\phi_k)\right], 
\end{align}
where we recall that $\psi_k$ is a real solution to Eq. \eqref{ermakov} and $\phi_{k}'=\psi_k^{-2}$. This last equality guarantees that $\phi_k$ grows monotonically in time, so that the sine and cosine functions appearing in this formula oscillate in time, generally in a $k$-dependent way. The overall oscillatory character of their contribution to $p_k$ depends on how fast they oscillate when compared to the relative variation of the global factor $\psi_k^2$. If they do vary faster, then they generally give rise to an oscillatory $p_k$ unless we have $A=B=1$ (or at least that these constants take values in a small neighbourhood of $1$), case in which the two coefficients of the sinusoidal terms are zero (or negligible). We recall that this can only happen for intervals of time and wavenumbers $k$ such that $k^2+s>0$, since we have seen that for $k^2+s\leq 0$ the function $p_k$ cannot oscillate owing to the dynamical equations \eqref{imre1} and \eqref{imre2}. It follows that we can characterize the NO spectra if we can restrict our considerations to real solutions of the Ermakov-Pinney equation such that, for $k^2+s>0$,
\begin{align}\label{nocond}
\left |\psi_{k}'\psi_{k}\right |< 1
\end{align}
(or much smaller than 1, if preferred), and to constants $A$ and $B$ in Eq. \eqref{generalp} that take values in a small neighbourhood of $1$. We have taken into account that the frequency of the sinusoidal functions in Eq. \eqref{generalp}, and therefore the rate at which they oscillate, is determined by $2\phi_{k}'=2\psi_k^{-2}$.

In order to analyze condition \eqref{nocond}, let us recall that Eq. \eqref{powereq} [equivalent to Eq. \eqref{ermakov} for real solutions $\rho_k$], is satisfied by every $p_k=|\mu_k|^2$, where $\mu_k$ is any normalized solution to Eq. \eqref{MS}. Conversely, any real $|\mu_k|$ such that its square satisfies Eq. \eqref{powereq} univocally leads to a normalized solution to Eq. \eqref{MS} [in virtue of Eq. \eqref{normalization}]. It then follows that, up to a sign, we can specify any particular real solution $\psi_k$ of the Ermakov-Pinney equation \eqref{ermakov} as $\sqrt{2}|\mu_k|$, where $|\mu_k|$ is completely determined by Eq. \eqref{muh} and $h_k$ is a solution to Eq. \eqref{diag} with strictly negative imaginary part. Therefore, we can rewrite the NO condition \eqref{nocond} in terms of $h_k$ as
\begin{align}\label{nocondh}
\left |\frac{\text{Re}(h_k)}{\text{Im}(h_k)}\right |< 1,
\end{align}
where we have used Eq. \eqref{imre2}.

For illustrative purposes, let us see whether the NO condition \eqref{nocondh} derived above is satisfied for fields with modes that obey Eq. \eqref{MS} in two situations where a natural choice of initial conditions is available. The first of these is when the mass $s$ is exactly a constant, namely $s'=0$, case in which Eq. \eqref{MS} represents the dynamical equation of a massive Klein-Gordon field in Minkowski spacetime. Natural initial conditions are then given by those corresponding to the Poincar\'e vacuum state for the field, with associated normalized solutions \eqref{muh} characterized by
\begin{align}
\text{Im}(h_k)=-\sqrt{k^2+s},
\end{align}
that is a constant. We see that the real part of $h_k$ is identically zero, and the NO condition indeed is satisfied for all $k$. In fact, from the general formula \eqref{generalp}, it follows that the Poincar\'e vacuum is the unique vacuum state for which the power spectrum is a constant, and therefore has the minimal oscillatory character. The second situation that we want to analyze is when the field $\mathcal{V}(\eta,\vec{x})$ describes the Mukhanov-Sasaki or the tensor perturbations of a cosmological background which is the de Sitter solution of General Relativity, in flat slicing. In this case, the mass $s$ coincides for both types of perturbations. It is given by $-2\eta^{-2}$, where the conformal time only takes negative values. Standard initial conditions for the normalized solutions to Eq. \eqref{MS} are in this case those specifying the Bunch-Davies vacuum state. These conditions give rise to solutions with \cite{mukhanov1}
\begin{align}
\text{Im}(h_k)=-\frac{k^3\eta^2}{1+\eta^{2}k^{2}}.
\end{align}
Using Eq. \eqref{imre2}, it follows that the real part of $h_k$ is given by
\begin{align}\label{reds}
\text{Re}(h_k)=\frac{1}{\eta+\eta^3 k^2}.
\end{align}
The NO condition is therefore satisfied when $k>|\eta^{-1}|$. It is worth noting that this inequality can always be satisfied for any $k$ by considering sufficiently large negative times, something that is certainly met in the limit $\eta\rightarrow -\infty$. Moreover, from Eq. \eqref{reds} for the real part of $h_k$, we see that the Bunch-Davies spectrum is completely monotonic in time for any $k$, so it does not display any oscillations. On the other hand, we know that the possible oscillatory behavior of any other power spectrum in de Sitter, that can be obtained by means of formula \eqref{generalp} setting $\psi_k$ as the solution selected by the Bunch-Davies conditions, must stop when $k^2+s\leq 0$. Since $s=-2\eta^{-2}$, this means that there are no oscillatory $p_k$ in de Sitter for $k\leq \sqrt{2}|\eta^{-1}|$ (again, when the conformal time tends to minus infinity, the restriction on $k$ disappears). Hence, we conclude that a NO spectrum for the Mukhanov-Sasaki or the tensor perturbations in a de Sitter background is obtained with the choice of a Bunch-Davies vacuum state, and any other NO spectrum must be in a small neighbourhood of it for $k>\sqrt{2}|\eta^{-1}|$, in the sense of setting the constants $A$ and $B$ close to $1$ in formula \eqref{generalp}.

\subsection{NO condition in effective hybrid LQC}

We have seen that the NO condition \eqref{nocondh} on the power spectrum is satisfied by the natural Poincar\'e and Bunch-Davies vacua on their respective Minkowski and de Sitter backgrounds. The main purpose of this section is to analyze if the condition can also be fulfilled in scenarios where the mass $s$ is given by effective hybrid LQC, at least in regimes where $k^2+s>0$ for wavenumbers in the phenomenological window $k\in[10^{-6},10^2]$ that covers, 
with some margin, the range corresponding to scales that we can consider observable nowadays, as we have commented \cite{Ivan,hybr-pred}. Actually, those regimes include the bounce, which can be understood as a privileged moment to set initial data. In particular, we are going to impose the NO condition at the time $\eta_0$ when the bounce occurs, and then study its stability throughout the period elapsed until the onset of inflation. In the case of dressed metric LQC, for a considerable part of the phenomenological window of wavenumbers $k$ that we are investigating, we have that $k^2+s\leq 0$ at the bounce owing to the negativity of the mass, that besides takes an absolute value of approximate order $10$ in Planck units \cite{mass}. Therefore, it seems unclear whether it is useful to impose NO conditions at the bouncing time in order to restrict the physically viable data in this case, at least as we have posed them; rather, one would have to appeal now to some \emph{additional} criteria to pick out the vacuum state, that then should satisfy the non-trivial requirement of leading to a suppression of the oscillations in the later Einsteinian period of kinetically dominated evolution of the background.

Let us first impose the NO condition \eqref{nocondh} at the time $\eta_0$ of the bounce, chosen as the moment to specify the Cauchy data of the normalized solutions to Eq. \eqref{MS}. This restricts the real part of $h_k$ at that time to be small compared to the imaginary part, so that
\begin{align}\label{nocondre}
|\text{Re}(h_k)(\eta_0) |= \epsilon_k|\text{Im}(h_k)(\eta_0)|,
\end{align}
where $\epsilon_k$ is a positive real number smaller than one\footnote{\label{prefer}Or much smaller than one, if preferred.}. For this restriction to hold in a small neighbourhood of $\eta_0$, it is necessary that the derivative of $\text{Re}(h_k)\text{Im}(h_k)^{-1}$ is also small initially, namely using Eqs. \eqref{imre1} and \eqref{imre2},
\begin{align}\label{nocondh2}
\left |\frac{k^2+s(\eta_0)}{\text{Im}(h_k)(\eta_0)}-(1+\epsilon_k^2)\text{Im}(h_k)(\eta_0) \right |< 1.
\end{align}
We note that, in general, $\epsilon_k^2$ provides a subdominant contribution to the second summand in this inequality. Besides, we recall that the mass $s$ in the kinetically dominated period (that includes the bounce) is given by the first equality in Eq. \eqref{hdmass} for hybrid LQC. Taking into account that the energy density at the bounce is a universal quantity, with a fixed value that is approximately a 41 percent of the Planck density, we have that $k^2+s(\eta_0)$ is always larger than one in Planck units. It then follows that a necessary condition for Eq. \eqref{nocondh} to hold is
\begin{align}\label{nocondim}
\text{Im}(h_k)(\eta_0)=-\sqrt{k^2+s(\eta_0)}+\delta_k, \qquad \frac{|\delta_k|}{\sqrt{k^2+s(\eta_0)}}<1
\end{align}
(again, see footnote \ref{prefer}) and we have used that the imaginary part of $h_k$ must be negative in order to provide normalized solutions to Eq. \eqref{MS}. Clearly, one can then choose the parameters $\epsilon_k$ and $\delta_k$ in such a way that the condition \eqref{nocondh2} on the derivative of $\text{Re}(h_k)\text{Im}(h_k)^{-1}$ is satisfied. Hence, in hybrid LQC, the NO condition can be guaranteed to hold in a small neighbourhood around the bounce if the initial data for the normalized solutions is constrained by Eqs. \eqref{nocondre} and \eqref{nocondim}, with sufficiently small parameters $\epsilon_k$ and $\delta_k$.

Let us now proceed to analyze the stability of the NO condition in the kinetically dominated regime that goes from the bounce to the onset of inflation. We recall that the dynamics experienced by the real and imaginary parts of $h_k$ are governed by a coupled set of real first order differential equations, or equivalently by a decoupled second order equation. Therefore, if the derived conditions \eqref{nocondre} and \eqref{nocondim} on those functions are imposed initally, their stability under evolution over the interval of time where one wishes to eliminate oscillations is controlled by the second derivative of $\text{Re}(h_k)\text{Im}(h_k)^{-1}$, that hence must be small in absolute value. Using Eqs. \eqref{imre1} and \eqref{imre2} again, we must have, let us say,
\begin{align}\label{stability}
|s'-4(k^2+s)\text{Re}(h_k)|< |\text{Im}(h_k)|.
\end{align}
We recall now that the energy density depends on time as $\rho\propto a^{-6}$ in the kinetically dominated region that includes the bounce. Since then $s'\propto a'$, the time derivative of the mass is zero at the bounce and the above stability condition reduces there to
\begin{align}
\epsilon_k< \left |\frac{1}{4[k^2+s(\eta_0)]}\right |,
\end{align}
which is perfectly compatible with our previous restrictions to guarantee NO power spectra in a small neighbourhood of the bounce in hybrid LQC.

In order to check the consistency of our conditions deeper into the kinetically dominated regime, we need more details about the behavior of $s'$ there. From the first equality in Eq. \eqref{hdmass} and the behavior $\rho\propto a^{-6}$ of the energy density, we straightforwardly see that $s'$ must be negative throughout this whole period (see also Fig. \ref{fig1}). Furthermore, we recall that in this region $aH=a'/a$ reaches only one maximum of order one, in Planck units, after the superinflationary regime that follows the bounce. Since the scale factor remains almost constant during superinflation, we have that $|s'|$ reaches only one maximum during the kinetically dominated period, that turns out to be approximately four times bigger than $s(\eta_0)$. Afterwards, $|s'|$ rapidly decreases to negligible values after roughly $1$ or $2$ e-folds (as it is confirmed in Fig. \ref{fig1}). Therefore, if we want a behavior of the form \eqref{nocondre} and \eqref{nocondh2} for the real and imaginary parts of $h_k$ that guarantees the NO condition at times after the bounce in the kinetically dominated regime, the only possible tension with stability, governed by Eq. \eqref{stability}, may arise in the region around the end of superinflation. Indeed, elsewhere we have that $s'$ contributes negligibly to this inequality, which is then compatible with the NO condition in a similar way as it was at the bounce. Actually, given that $4(k^2+s)$ is of the order of $|s'|$ or larger around the end of superinflation, we think it is likely that this tension disappears if one chooses properly the initial parameters $\epsilon_k$ and $\delta_k$.

We conclude that, in effective hybrid LQC, conditions for NO power spectra on the gauge invariant perturbations can be consistently set at the bounce via Eqs. \eqref{nocondre} and \eqref{nocondim} with appropriately small parameters $\epsilon_k$ and $\delta_k$, that in principle can be chosen without obstructions. With such a suitable choice, the associated spectra should display a stable NO behavior throughout the kinetically dominated period after the bounce. Moreover, among the normalized solutions to Eq. \eqref{MS} of the form \eqref{muh} restricted by these NO considerations, we notice that there consistently exist some that satisfy conditions \eqref{constsol} and \eqref{constsol2} at the beginning of the classical, Einsteinian kinetically dominated regime. These would lead to power spectra that remain approximately constant throughout this classical period so that, even when evaluated around the onset of inflation, they can still provide useful information about the two-point function of the perturbations at those primeval stages right when the effective LQC corrections became negligible.

\section{Uniqueness of the NO spectrum in the ultraviolet regime}

The procedure of Hamiltonian diagonalization carried out in Section II, using explicitly time-dependent transformations, parallels a similar construction performed in Ref. \cite{msdiag} for the fully canonical formulation of the classical system formed by a homogeneous FLRW background with perturbations, truncated at lowest non-trivial order in the action. That work addresses the possibility of diagonalizing the resulting (quadratic) perturbative contribution of gauge invariants to the zero mode of the Hamiltonian constraint of the full system, employing transformations of the form \eqref{newvar} and \eqref{sympl} with coefficients that depend on the canonical variables which describe the homogeneous background cosmology. If one completes these transformations to be canonical in the entire system, the perturbative contributions to the zero mode of the Hamiltonian constraint turn out to be precisely of the form \eqref{newham} (up to a global factor $a^{-1}$ for a standard lapse), where the time derivatives are replaced by conformal Poisson brackets with the Hamiltonian of the unperturbed FLRW cosmology \cite{msdiag}. These perturbative contributions are then diagonal in this context if $h_k$ satisfies Eq. \eqref{diag}, after replacing the time derivative by the mentioned Poisson brackets. 

Focusing on the asymptotic regime of unboundedly large wavenumbers $k$, or ultraviolet regime, it was shown in Ref. \cite{msdiag} that it is possible to eliminate each contribution to the non-diagonal terms in the perturbative part of the Hamiltonian constraint, order by order in powers of $k$. The result is an ultraviolet diagonalization characterized by a very specific asymptotic expansion of at least one solution $h_k$ to the analog of our Eq. \eqref{diag} in that work. Following a completely similar procedure, we can perform the same type of asymptotic diagonalization of our Hamiltonian \eqref{newham}, in the regime of large $k$, yielding the expansion \cite{msdiag}
\begin{align}\label{hasymp}
kh_k^{-1}\sim i\left[1-\frac{1}{2k^2}\sum_{n=0}^{\infty}\left(\frac{-i}{2k}\right)^{n}\gamma_n \right],
\end{align}
where the coefficients $\gamma_n$ are real, only depend on time, and are given by the following iterative relation, that is deterministic:
\begin{align}\label{gamman}
\gamma_{0}=s,\qquad \gamma_{n+1}=-\gamma_{n}'+4s\left[\gamma_{n-1}+\sum_{m=0}^{n-3}\gamma_m \gamma_{n-(m+3)}\right]-\sum_{m=0}^{n-1}\gamma_m \gamma_{n-(m+1)}.
\end{align}
We define $\gamma_{-n}=0$ for all $n>0$. This leads to a unique asymptotic expansion of, at least, one solution $h_k$ to Eq. \eqref{diag}, with imaginary part that is strictly negative \cite{msdiag}. Therefore, it provides in turn a very precise asymptotic expansion of, at least, one normalized solution to Eq. \eqref{MS}, via Eq. \eqref{muh}. We call any such solution $\tilde{\mu}_k$. Its associated square norm $\tilde{p}_k=|\tilde{\mu}_k|^2$ is then of the form
\begin{align}
\tilde{p}_k=\frac{1}{2k}\left(1-\Gamma_k\right),
\end{align}
where $\Gamma_k$ has the following asymptotic expansion:
\begin{align}\label{uvgamma}
\Gamma_k \sim 
\frac{1}{2k^2}\left[1-\frac{1}{2k^2}\sum_{n=0}^{\infty}\bigg(\frac{i}{2k}\bigg)^{2n}\gamma_{2n} \right]^{-1}\sum_{n=0}^{\infty}\bigg(\frac{i}{2k}\bigg)^{2n}\left[\gamma_{2n}-\frac{1}{2k^2}\sum_{m=0}^{2n}(-1)^{m}\gamma_m \gamma_{2n-m}\right].
\end{align}
This is a series where each summand depends on the wavenumber $k$ only through an even inverse power of it. Since we know that any such $\tilde{p}_k$ must be of the form $\tilde{\psi}_k^2 /2$, where $\tilde{\psi}_k$ is a real solution to the Ermakov-Pinney equation \eqref{ermakov}, this procedure of ultraviolet diagonalization fixes as well (up to sign) a very specific asymptotic expansion of, at least, one solution $\tilde{\psi}_k$ to that equation, in the regime of unboundedly large $k$. Let us precisely take it (or one of them, if there were more than one) as the particular solution to insert in the general formula \eqref{generalp} for any other power spectrum. Then, any function $p_k$ equal to the square norm of a normalized solution to Eq. \eqref{MS} is given by
\begin{align}\label{uvp}
p_k = \frac{1}{4k}\left(1-\Gamma_k\right)\left[A+B+ (A-B)\cos(2k\eta+2\theta_k)+2C\sin(2k\eta+2\theta_k)\right],
\end{align}
where $\Gamma_k$ has the asymptotic expansion \eqref{uvgamma} and $\theta_k$ is a function with dominant contribution in the ultraviolet regime of order $k^{-1}$. In this asymptotic regime, the dominant term in $p_k$ is
\begin{align}
\frac{1}{4k}\left[A+B+ (A-B)\cos(2k\eta+2\theta_k)+2C\sin(2k\eta+2\theta_k)\right],
\end{align}
because $\Gamma_k$ is of order $k^{-2}$. This is a highly oscillatory function for unboundedly large $k$ unless we strictly impose $A=B=1$, in which case $p_k$ reduces to $\tilde{p}_k$. Actually, this choice of constants is the only one that succeeds to eliminate, order by order in the expansion of $\Gamma_k$ in inverse powers of $k$, all the scale-dependent oscillations in the considered ultraviolet regime. 

In this way, we conclude that, in the asymptotic regime of unboundedly large wavenumbers (or short wavelength scales) there exists only one expansion for the power spectrum of the field $\mathcal{V}(\eta,\vec{x})$ for which absolutely no $k$-dependent oscillations occur over time. Let us notice that this result holds for any mass $s$ that is a smooth function of time. Moreover, any choice of normalized solutions to Eq. \eqref{MS} that presents a completely NO spectrum for all $k$, if such a choice exists, must have the asymptotic expansion characterized by Eqs. \eqref{hasymp} and \eqref{gamman}. It is reasonable to expect that there exists at least one set of normalized solutions that are, e.g., continuous functions of $k$ and possess such asymptotic expansion. If such continuity in $k$ ensures that the NO condition in the previous section is satisfied for all scales, then we would have a set of preferred choices of power spectrum for the field $\mathcal{V}(\eta,\vec{x})$. There is actual evidence that this should be the case in the context of effective LQC, just by looking at the properties of the asymptotic expansion. Indeed, it has been shown that when $s$ corresponds to a constant or to the mass for cosmological perturbations on a de Sitter background, the series given by Eqs. \eqref{hasymp} and \eqref{gamman} converges for sufficiently large $k$ \cite{msdiag}. The resulting functions are analytically well defined at all other scales, and they correspond to the choices of normalized solutions set by the Poincar\'e and Bunch-Davies vacuum states, respectively \cite{msdiag}. As we explicitly saw in the previous section, these two states lead to power spectra that display no oscillations at all. Furthermore, the Einsteinian regime reached as part of the effective LQC evolution of the primordial Universe presents two regions where $s$ is very close to either a constant or to the mass for a de Sitter background: these are the classical kinetically dominated period and the slow-roll inflationary phase, respectively. It therefore seems reasonable to expect that there should exist at least one choice of normalized solutions to Eq. \eqref{MS}  that is continuous in $k$ and has an asymptotic expansion fixed by Eqs. \eqref{hasymp} and \eqref{gamman} {\emph {at all finite times}} such that it gives rise to power spectra that are of NO type in the regime where effective LQC reproduces the classical FLRW evolution. Any such choice would, in turn, correspond to a promising candidate for the vacuum state at the bounce, with a power spectrum capable of capturing the traces left by LQC effects in the dynamics of the primordial perturbations before these effects became ignorable in the background evolution.

\section{Conclusions}

We have investigated from a theoretical point of view the possibility of obtaining non-oscillating power spectra for primordial perturbations in cosmology, putting a special emphasis on cosmological backgrounds that correspond to certain solutions of effective LQC with inflation, such that they display a pre-inflationary regime that is dominated by the kinetic energy density of the inflaton. This type of background is phenomenologically favoured when confronting the expected loop quantum geometry effects on the evolution of the perturbations with CMB observations. Furthermore, we have characterized the general conditions that any power spectrum must satisfy in order to eliminate or minimize its scale-dependent oscillations over time, making use of a well-known equivalence between our hyperbolic field equations with a time-dependent mass and the Ermakov-Pinney equation. Finally, we have discussed the uniqueness of the NO power spectrum in the ultraviolet regime of short wavelength scales, concluding that there is only one asymptotic expansion that displays no scale-dependent oscillations at all. This expansion actually corresponds to the choice of a standard Poincar\'e or a Bunch-Davies vacuum, respectively, for a Minkowski or a de Sitter background, and constitutes a promising line of attack to completely fix the initial conditions for the primordial perturbations in effective LQC by means of a physically well-motivated criterion.

In more detail, we have first considered the general equation of a harmonic oscillator with a time-dependent mass, and have conveniently characterized its normalized solutions by diagonalizing the associated Hamiltonian employing explicitly time-dependent transformations. This is the type of equation that each mode of the gauge invariant perturbations satisfies, not only in classical perturbation theory around FLRW cosmology, but also in the context of the hybrid and dressed metric approaches to LQC when the unperturbed cosmology can be described effectively. Then, using general features of the effective LQC backgrounds of interest and of the solutions to the considered harmonic oscillator equation, we have discussed the qualitative impact that oscillatory power spectra may have on observations. We have argued that, in order to get rid of any net amplification of power artificially pumped by oscillations in classical regimes where the classical cosmological evolution is recovered, as well as to obtain a neat information about the quantum state of the perturbations in stages where the LQC modifications may not yet be completely negligible in the background evolution, we need to focus our attention on initial conditions that lead to NO spectra.

We have then studied the general conditions that the normalized solutions to our field equation must satisfy in order to avoid the presence of scale-dependent oscillations over time in their associated spectra. For that, we have written any possible power spectrum in terms of one particular solution to the Ermakov-Pinney equation that corresponds to our hyperbolic equations with time-dependent mass, in a way that makes manifest the possible oscillations. Imposing that these oscillations have a minimal contribution in the admissible power spectra results into a very specific condition on the particular solution to the Ermakov-Pinney equation and on the two integration constants that fix each power spectrum in terms of it. We have analyzed if this NO condition can be consistently imposed at the bounce that replaces the classical cosmological singularity, for perturbations in hybrid LQC. The result is in the afirmative. We have also checked that there are no serious obstructions to extend this requirement from the bounce all the way to the onset of inflation. On the other hand, in the case of the dressed metric approach to LQC, we have argued that there is no clear motivation from our analytical considerations to substantiate the imposition of the NO condition at the bounce for the scales of observational interest, owing to the fact that the associated negativity of the time-dependent mass around the bounce implies that, in this case, the oscillations at the considered scales can start only in a later phase of the evolution, which is actually when the main LQC effects are negligible. This fact leads to the need of additional criteria or extra input in order to pick out the initial conditions at the bounce in the dressed metric formalism. Nonetheless, these criteria or input should be non-trivially constrained by requiring the NO condition in the part of the evolution where the background reaches the classical, Einsteinian regime.

To conclude our analysis, we have investigated the asymptotic behavior of the power spectrum in the sector of unboundedly large wavenumbers $k$. Taking insight from previous results about asymptotic Hamiltonian diagonalization for cosmological perturbations \cite{msdiag}, we have determined one specific asymptotic behavior for certain solutions to the Ermakov-Pinney equation, given as a series in inverse powers of $k$. We have then inserted this asymptotic series in the general formula for power spectra previously derived. The resulting expression manifestly displays rapid oscillations at every order in inverse powers of $k$, except for a single choice of the otherwise free integration constants. This allows us to conclude that there is only one possible asymptotic NO behavior for the power spectrum, that we have completely characterized. We have finally argued how this asymptotic expansion can reasonably lead, by imposing continuity in the scale $k$, to a unique (set of) choice(s) of normalized solutions to our field equations with a power spectrum that satisfies the NO condition for all $k$, in the entire classical pre-inflationary and inflationary phases of effective LQC. 
The choice suggested by this procedure constitutes a promising candidate as a physically distinguished vacuum state for the cosmological perturbations in effective LQC.

Our work provides an important step towards the analytical characterization of a reasonable set of initial conditions for the cosmological fluctuations in a pre-inflationary Universe with LQC effects. Completing the specification of these data would not only confer more robustness to the predictions that can be drawn from approaches to LQC such as the hybrid or the dressed metric approaches (as well as to allow one to clearly isolate those predictions from other classical effects, like e.g. the ones arising from a short-lived inflation). Actually, it would be a key ingredient to understand the consequences of these various theoretical models in an analytical way, and discriminate between them. Moreover, it would allow one to falsify them against the CMB observations without the shadow that quantum field theory ambiguities cast on such possible tests nowadays. 

Finally, it is worth noticing that the conditions found here for NO power spectra have been obtained for general and unspecified time-dependent (differentiable) mass functions of the perturbations. The same is true for our characterization of a unique asymptotic expansion for such spectra. In this respect, our analysis potentially serves as a first contribution to the study of preferred choices of a vacuum for primordial perturbations in other theoretical approaches to cosmology apart from LQC (e.g. in the context of bouncing cosmologies \cite{bouncing}), that produce modifications to the time-dependent mass of the perturbations with respect to its behavior in the standard inflationary paradigm.

\acknowledgments

This work was supported by Project. No. MINECO FIS2017-86497-C2-2-P from Spain. The authors would especially like to thank Daniel Mart\'{\i}n de Blas and Javier Olmedo for discussions regarding the value of non-oscillating spectra in effective Loop Quantum Cosmology.


\begin{thebibliography}{299}

\bibitem{planck} P.A.R. Ade {\it et al.} (Planck Collaboration), Planck 2015 results. XIII. Cosmological parameters, A\&A {\bf 594}, A13 (2016).

\bibitem{planck-inf} P.A.R. Ade {\it et al.} (Planck Collaboration). Planck 2015 results. XX. Constraints on inflation, A\&A {\bf 594}, A20 (2016).

\bibitem{structures} A.R. Liddle and D.H. Lyth, {\it Cosmological Inflation and Large-Scale Structure} (Cambridge University Press, Cambridge, England, 2000).

\bibitem{mukhanov1} V. Mukhanov, {\it Physical Foundations of Cosmology} (Cambridge University Press, Cambridge, England, 2005).

\bibitem{HalliwellHawking}  J.J. Halliwell and S.W. Hawking, Origin of structure in the Universe, Phys. Rev. D {\bf 31}, 1777 (1985).

\bibitem{pintoneto1} E.J.C. Pinho and N. Pinto-Neto, Scalar and vector perturbations in quantum cosmological backgrounds, Phys. Rev. D {\bf 76}, 023506 (2007).

\bibitem{pintoneto2} F.T. Falciano and N. Pinto-Neto, Scalar perturbations in scalar field quantum cosmology, Phys. Rev. D {\bf 79}, 023507 (2009).

\bibitem{AshLewaDress} A. Ashtekar, W. Kaminski, and J. Lewandowski, Quantum field theory on a cosmological, quantum space-time, Phys. Rev. D {\bf 79}, 064030 (2009).

\bibitem{Kiefer1} C. Kiefer and M. Kr{\"a}mer, Quantum gravitational contributions to the cosmic microwave background anisotropy spectrum, Phys. Rev. Lett. {\bf 108}, 021301 (2012).

\bibitem{Bojo0} M. Bojowald, G.M. Hossain, M. Kagan, and S. Shankaranarayanan, Anomaly freedom in perturbative loop quantum gravity, Phys. Rev. D {\bf 78}, 063547 (2008).

\bibitem{Bojo1} M. Bojowald, G. Calcagni, and S. Tsujikawa, Observational constraints on loop quantum cosmology, Phys. Rev. Lett. {\bf 107}, 211302 (2011).

\bibitem{CLB} T. Cailleteau, L. Linsefors, and A. Barreau, Anomaly-free perturbations with inverse-volume and holonomy corrections in loop quantum cosmology, Classical Quantum Gravity {\bf 31}, 125011 (2014).

\bibitem{Bojo2} A. Barrau, M. Bojowald, G. Calcagni, J. Grain, and M. Kagan, Anomaly-free cosmological perturbations in effective canonical quantum gravity, JCAP {\bf 05} (2015) 051.

\bibitem{Edward} E. Wilson-Ewing, Testing loop quantum cosmology, Comptes Rendus Physique {\bf 18}, 207 (2017).

\bibitem{Edward2}  F. Gerhardt, D. Oriti, and E. Wilson-Ewing, The separate universe framework in group field theory condensate cosmology, Phys. Rev. D {\bf 98}, 066011 (2018). 

\bibitem{alesci1} E. Alesci, A. Barrau, G. Botta, K. Martineau, and G. Stagno, Phenomenology of quantum reduced loop gravity in the isotropic cosmological sector, Phys. Rev. D {\bf 98}, 106022 (2018).

\bibitem{alesci2} J. Olmedo and E. Alesci, Power spectrum of primordial perturbations for an emergent universe in quantum reduced loop gravity, JCAP {\bf 04} (2019) 030.

\bibitem{AAN3} I. Agullo, A. Ashtekar, and W. Nelson, A quantum gravity extension of the inflationary scenario, Phys. Rev. Lett. {\bf 109}, 251301 (2012).

\bibitem{AAN1}  I. Agullo, A. Ashtekar, and W. Nelson, Extension of the quantum theory of cosmological perturbations to the Planck era, Phys. Rev. D {\bf 87}, 043507 (2013).

\bibitem{AAN2}  I. Agullo, A. Ashtekar, and W. Nelson, The pre-inflationary dynamics of loop quantum cosmology: Confronting quantum gravity with observations, Classical Quantum Gravity {\bf 30}, 085014 (2013).

\bibitem{Ivan} I. Agullo and N.A. Morris, Detailed analysis of the predictions of loop quantum cosmology for the primordial power spectra, Phys. Rev. D {\bf 92}, 124040 (2015).

\bibitem{hyb-pert1} M. Fern\'andez-M\'endez, G.A. Mena Marug\'an, and J. Olmedo, Hybrid quantization of an inflationary universe, Phys. Rev. D {\bf 86}, 024003  (2012).

\bibitem{hyb-pert2}  M. Fern\'andez-M\'endez, G.A. Mena Marug\'an, and J. Olmedo, Hybrid quantization of an inflationary model: The flat case, Phys. Rev. D {\bf 88}, 044013 (2013).

\bibitem{hyb-pert-eff} M. Fern\'andez-M\'endez, G.A. Mena Marug\'an, and J. Olmedo, Effective dynamics of scalar perturbations in a flat Friedmann-Robertson-Walker spacetime in loop quantum cosmology, Phys. Rev. D {\bf 89}, 044041 (2014).

\bibitem{hyb-pert3} L. Castell\'o Gomar, M. Fern\'andez-M\'endez, G.A. Mena Marug\'an, and J. Olmedo, Cosmological perturbations in hybrid loop quantum cosmology: Mukhanov--Sasaki variables, Phys. Rev. D {\bf 90}, 064015 (2014).

\bibitem{hyb-pert4} L. Castell\'o Gomar, M. Mart\'{\i}n-Benito, and G.A. Mena Marug\'an, Gauge-invariant perturbations in hybrid quantum cosmology, JCAP {\bf 06} (2015) 045.

\bibitem{hyb-pert5} L. Castell\'o Gomar, M. Mart\'{\i}n-Benito, and G.A. Mena Marug\'an, Quantum corrections to the Mukhanov-Sasaki equations, Phys. Rev. D {\bf 93}, 104025 (2016).

\bibitem{hybr-ten} F. Ben\'{\i}tez Mart\'{\i}nez and J. Olmedo, Primordial tensor modes of the early universe, Phys. Rev. D {\bf 93}, 124008 (2016).

\bibitem{hybr-pred} L. Castell\'o Gomar, G.A. Mena Marug\'an, D. Mart\'{\i}n de Blas, and J. Olmedo, Hybrid loop quantum cosmology and predictions for the cosmic microwave background, Phys. Rev. D {\bf 96}, 103528 (2017).

\bibitem{bojo} M. Bojowald, Loop quantum cosmology, Living Rev. Rel. {\bf 11}, 4 (2008).

\bibitem{abl} A. Ashtekar, M. Bojowald, and J. Lewandowski, Mathematical structure of loop quantum cosmology, Adv. Theor. Math. Phys. {\bf 7}, 233 (2003).

\bibitem{ashparam} A. Ashtekar and P. Singh, Loop quantum cosmology: A status report, Classical Quantum Gravity {\bf 28}, 213001 (2011).

\bibitem{lqcmena} G.A. Mena Marug\'an, A brief introduction to loop quantum cosmology, AIP Conf. Proc. {\bf 1130}, 89  (2009).

\bibitem{AshNe} A. Ashtekar, B. Gupt, D. Jeong, and V. Sreenath, Alleviating the tension in CMB using Planck-scale physics, Phys. Rev. Lett. {\bf 125}, 051302 (2020).

\bibitem{APS} A. Ashtekar, T. Paw{\l}owski, and P. Singh, Quantum nature of the big bang, Phys. Rev. Lett. {\bf 96}, 141301 (2006).

\bibitem{mmo} M. Mart\'{\i}n-Benito, G.A. Mena Marug\'an, and J. Olmedo, Further improvements in the understanding of isotropic loop quantum cosmology, Phys. Rev. D {\bf 80}, 104015 (2009).

\bibitem{mass} B. Elizaga Navascu\'es, D. Mart\'{\i}n de Blas, and G.A. Mena Marug\'an, Time-dependent mass of cosmological perturbations in the hybrid and dressed metric approaches to loop quantum cosmology, Phys. Rev. D {\bf 97}, 043523 (2018).

\bibitem{BD} T.S. Bunch and P. Davies, Quantum field theory in de Sitter space: Renormalization by point splitting, Proc. R. Soc. Lond. A {\bf 360}, 117 (1978).

\bibitem{uniquenessflat} L. Castell\'o Gomar, J. Cortez, D. Mart\'{\i}n-de Blas, G.A. Mena Marug\'an, and J.M. Velhinho, Uniqueness of the Fock quantization of scalar fields in spatially flat cosmological spacetimes, JCAP {\bf 11} (2012) 001.

\bibitem{uniquenessrep} J. Cortez, G.A. Mena Marug\'an, J. Olmedo, and J.M. Velhinho, A uniqueness criterion for the Fock quantization of scalar fields with time-dependent mass, Classical Quantum Gravity {\bf 28}, 172001 (2011).

\bibitem{adiabatic1} L. Parker, Quantized fields and particle creation in expanding universes. I, Phys. Rev. {\bf 183}, 1057 (1969).

\bibitem{adiabatic2} C. L{\"u}ders and J.E. Roberts, Local quasiequivalence and adiabatic vacuum states, Commun. Math. Phys. {\bf 134}, 29 (1990).

\bibitem{AGvacio1} A. Ashtekar and B. Gupt, Initial conditions for cosmological perturbations, Classical Quantum Gravity {\bf 34}, 035004 (2017).

\bibitem{AGvacio2} A. Ashtekar and B. Gupt, Quantum gravity in the sky: Interplay between fundamental theory and observations, Classical Quantum Gravity {\bf 34}, 014002 (2017).

\bibitem{Universe} B. Elizaga Navascu\'es, D. Mart\'{\i}n de Blas, and G.A. Mena Marug\'an, The vacuum state of primordial fluctuations in hybrid loop quantum cosmology, Universe {\bf 4}, 98 (2018).

\bibitem{no} D. Mart\'{\i}n de Blas and J. Olmedo, Primordial power spectra for scalar perturbations in loop quantum cosmology, JCAP {\bf 06} (2016) 029.

\bibitem{ermakov} V.P. Ermakov, Second-order differential equations. Conditions of complete integrability, Univ. Izv. Kiev {\bf 20}, 1 (1880).

\bibitem{pinney} E. Pinney, The nonlinear differential equation $y (x) + p(x)y + cy^{-3} = 0$, Proc. Am. Math. Soc. {\bf 1}, 681 (1950).

\bibitem{MukhanovSasaki} V. Mukhanov, Quantum theory of gauge-invariant cosmological perturbations, Zh. Eksp. Teor. Fiz. {\bf 94}, 1 (1988) [Sov. Phys. JETP {\bf 67}, 1297 (1988)].

\bibitem{sasaki} M. Sasaki,  Gauge invariant scalar perturbations in the new inflationary universe, Prog. Theor. Phys. {\bf 70}, 394 (1983).

\bibitem{sasakikodama} H. Kodama and M. Sasaki, Cosmological perturbation theory, Prog. Theor. Phys. Suppl. {\bf 78}, 1 (1984).

\bibitem{langlois} D. Langlois, Inflation and cosmological perturbations, Lect. Notes Phys. {\bf 800}, 1 (2010).

\bibitem{fermiLQC} B. Elizaga Navascu\'es, M. Mart\'{\i}n-Benito, and G.A. Mena Marug\'{a}n, Fermions in hybrid loop quantum cosmology, Phys. Rev. D {\bf 96}, 044023 (2017).

\bibitem{fermiback} B. Elizaga Navascu\'es, G.A. Mena Marug\'an, and S. Prado Loy, Backreaction of fermionic perturbations in the Hamiltonian of hybrid loop quantum cosmology, Phys. Rev. D {\bf 98}, 063535 (2018).

\bibitem{continuum} B. Elizaga Navascu\'es and G.A. Mena Marug\'an, Perturbations in hybrid loop quantum cosmology: Continuum limit in Fourier space, Phys. Rev. D {\bf 98}, 103522 (2018).

\bibitem{linde} C.R. Contaldi, M. Peloso, L. Kofman, and A. Linde, Suppressing the lower multipoles in the CMB anisotropies, JCAP {\bf 07} (2003) 002.

\bibitem{Wang1} T. Zhu, A. Wang, K. Kirsten, G. Cleaver, and Q. Sheng, Universal features of quantum bounce in loop quantum cosmology, Phys. Lett. {\bf B773}, 196 (2017).

\bibitem{Wang2} T. Zhu, A. Wang, K. Kirsten, G. Cleaver, and Q. Sheng, Pre-inflationary universe in loop quantum cosmology, Phys. Rev. D {\bf 96}, 083520 (2017).

\bibitem{jorma} A. Bhardwaj, E.J. Copeland, and J. Louko, Inflation in loop quantum cosmology, Phys. Rev. D {\bf 99}, 063520 (2019).

\bibitem{IvanG} I. Agullo, Loop quantum cosmology, non-Gaussianity, and CMB power asymmetry, Phys. Rev. D {\bf 92}, 064038 (2015).

\bibitem{Ivan2} I. Agullo, D. Kranas, and V. Sreenath, Anomalies in the CMB from a cosmic bounce, arXiv:2005.01796 (2020).

\bibitem{it} C. Bertoni, F. Finelli, and G. Venturi, Adiabatic invariants and scalar fields in a de Sitter space-time, Phys. Lett. A {\bf 237}, 331 (1998).

\bibitem{hawk} R.M. Hawkins and J.E. Lidsey, Ermakov-Pinney equation in scalar field cosmologies, Phys. Rev. D {\bf 66}, 023523 (2002).

\bibitem{kam1} A.Y. Kamenshchik, A. Tronconi, and G. Venturi, Inflation and quantum gravity in a Born-Oppenheimer context, Phys. Lett. B {\bf 726}, 518 (2013).

\bibitem{giesel} M.J. Fahn, K. Giesel, and M. Kobler, Dynamical properties of the Mukhanov-Sasaki Hamiltonian in the context of adiabatic vacua and the Lewis-Riesenfeld invariant, Universe {\bf 5}, 170 (2019).

\bibitem{milne} W.E. Milne, The numerical determination of characteristic numbers, Phys. Rev. {\bf 35}, 863 (1930).

\bibitem{msdiag} B. Elizaga Navascu\'es, G.A. Mena Marug\'an, and T. Thiemann, Hamiltonian diagonalization in hybrid quantum cosmology, Classical Quantum Gravity {\bf 36}, 18 (2019).

\bibitem{bouncing} R. Brandenberger and P. Peter, Bouncing cosmologies: Progress and problems, Found. Phys. {\bf 47}, 797 (2017).


\end{thebibliography}
\end{document}